\documentclass[twocolumn]{aastex61}
\usepackage{natbib}

\newcommand{\logg}{\ensuremath{\log g}}

\newcommand\aastex{AAS\TeX}

\received{\today}
\shorttitle{\aastex\ A mega metal-poor dwarf star}
\shortauthors{Aguado et al.}
\begin{document}
   \title{J0023+0307: a Mega Metal-poor Dwarf Star from SDSS/BOSS\footnote{
Based on observations made with William Herschel Telescope (WHT) and the Gran Telescopio de Canarias (GTC), at the Observatorio del Roque de los Muchachos of the Instituto de Astrof\'{\i}sica de Canarias, in La Palma.}
 }

\correspondingauthor{David~S. Aguado}
%\email{aguado@iac.es, jonay@iac.es, callende@iac.es, rrl@iac.es}
\email{aguado@iac.es}

%\author{IAC EMP GROUP}
\author{David~S. Aguado}
\affil{Instituto de Astrof\'{\i}sica de Canarias,
            V\'{\i}a L\'actea, 38205 La Laguna, Tenerife, Spain\\}
\affiliation{Universidad de La Laguna, Departamento de Astrof\'{\i}sica, 
             38206 La Laguna, Tenerife, Spain \\}             

\author{ Carlos Allende Prieto}
\affil{Instituto de Astrof\'{\i}sica de Canarias,
              V\'{\i}a L\'actea, 38205 La Laguna, Tenerife, Spain\\}
 \affiliation{Universidad de La Laguna, Departamento de Astrof\'{\i}sica, 
             38206 La Laguna, Tenerife, Spain \\}             
\author{Jonay~I. Gonz\'alez Hern\'andez}
\affil{Instituto de Astrof\'{\i}sica de Canarias,
              V\'{\i}a L\'actea, 38205 La Laguna, Tenerife, Spain\\}
 \affiliation{Universidad de La Laguna, Departamento de Astrof\'{\i}sica, 
             38206 La Laguna, Tenerife, Spain \\} 
\author{Rafael Rebolo}
\affil{Instituto de Astrof\'{\i}sica de Canarias,
              V\'{\i}a L\'actea, 38205 La Laguna, Tenerife, Spain\\}
 \affiliation{Universidad de La Laguna, Departamento de Astrof\'{\i}sica, 
             38206 La Laguna, Tenerife, Spain \\}             

\affiliation{ Consejo Superior de Investigaciones Cient\'{\i}ficas, 28006 Madrid, Spain\\}

\begin{abstract}
Only a handful of stars have been identified with an iron abundance [Fe/H]$<-5$, and only one at [Fe/H]$<-7$. These stars have very large carbon-to-iron ratios, with  A(C)$\sim 7.0$, most likely due to fallback in core-collapse supernovae, which makes their total metallicity $Z$ much higher than their iron abundances.  The failure to find population III stars, those with no metals, has been interpreted, with support from theoretical modeling, as the result of a top-heavy initial mass function. With zero or very low metal abundance limiting radiative cooling, the formation of low-mass stars could be inhibited. Currently, the star SDSS J1029$+$1729 sets the potential metallicity threshold for the formation of low-mass stars at $\log Z/Z_{\odot} \sim -5$.  We have identified  SDSS J0023+0307, a primitive star with $T_{\rm eff}=$6188$\pm 84$\,K, and \logg=4.9$\pm0.5$,  an upper limit [Fe/H]$<-6.6$, and a carbon abundance  A(C)$< 6.3$. In our quest to push down the metallicity threshold we find J0023+0307 to be one of the two most iron-poor stars known, and it exhibits less carbon that most of stars at [Fe/H]$<-5$.
\end{abstract}

\keywords{stars: PopulationII -- stars: abundances -- stars: PopulationIII -- Galaxy:abundances -- Galaxy:formation -- Galaxy:halo}

\section{Introduction} \label{intro}

The most interesting halo stars are those that formed in the first or second generation. Such objects are extremely rare: despite the efforts, only a few stars are known at [Fe/H]$<-5$. SMSS J0313--6708, the most iron-poor star known, was discovered a few years ago at [Fe/H]$<-7.1$ by the SkyMapper Southern Sky Survey \citep{kel14}.  More recently, \citet{bes15} derived [Fe/H]$<-7.5$ (3D, NLTE) and [Fe/H]$<-7.8$ (1D, LTE). The lack of metals in the gas available in the $\sim 10^6$ M$_{\odot}$ mini-haloes where the first stars formed, at $z \sim$ 20,  limits radiative cooling increasing the Jeans mass and shifting the IMF to large masses, to the point that perhaps no low-mass ($<$ 1M$_{\odot}$) stars were formed in the first generation (Bromm \& Loeb 2003). This picture has been challenged by one of these stars, SDSS J1029+1729 (Caffau et al. 2011), which not only shows [Fe/H]$\sim$-5, but also relatively low C and N abundances, suggesting that low mass stars can form even at such low metallicities. 

We have performed an extensive search for extremely metal-poor stars using low-resolution spectra from the Sloan Extension for Galactic Understanding and Exploration (SEGUE, \citealt{yan09}), the Baryonic Oscillations Spectroscopic Survey (BOSS, \citealt{eis11,daw13}) and the large Sky Area Multi-Object Fiber Spectroscopic Telescope (LAMOST, \citealt{deng12}). Complete details on the target selection can be found in \citet{agu17I,agu18I}. Two years ago we reported  on SDSS J1313-0019 \citep{alle15,fre15,agu17I}, a carbon-enhanced metal-poor red giant star with [Fe/H]$= -4.7$ and [C/Fe]$\sim +2.8$. More recently, we  identified SDSS J0815+4729, a dwarf star with [Fe/H]$\le -5.8$ and [C/Fe]$>+5$ \citep{agu18I}. 

We have now identified a new mega metal-poor dwarf with [Fe/H]$< -6.6$, J0023+0307, from  SDSS/BOSS spectra ($\lambda/\Delta \lambda\sim$2, 000). Follow-up spectroscopy with the 10.4m GTC and the 4.2m WHT telescopes confirms the metallicity  determination based on SDSS data. The spectrum of this star does not show the usual enhancement in carbon found in all other stars at [Fe/H]$<-5$, and has an obvious ISM contribution in the Ca II H \& K region, offset in velocity from the stellar lines. A stellar Ca II K line appears barely visible in the spectra, and could be affected by further ISM contributions. Therefore, our estimated iron abundance (based on the assumption that the star has a typical [Ca/Fe]$\sim$0.4), is only an upper limit, and since carbon does not appear enhanced at the levels found in other hyper metal-poor stars, J0023+0307 could well be the most metal-poor star known. Additional observations at higher resolution and signal-to-noise ratio (S/N) are required to confirm this is the case.

\begin{figure}
\begin{center}
{\includegraphics[width=90 mm, angle=0]{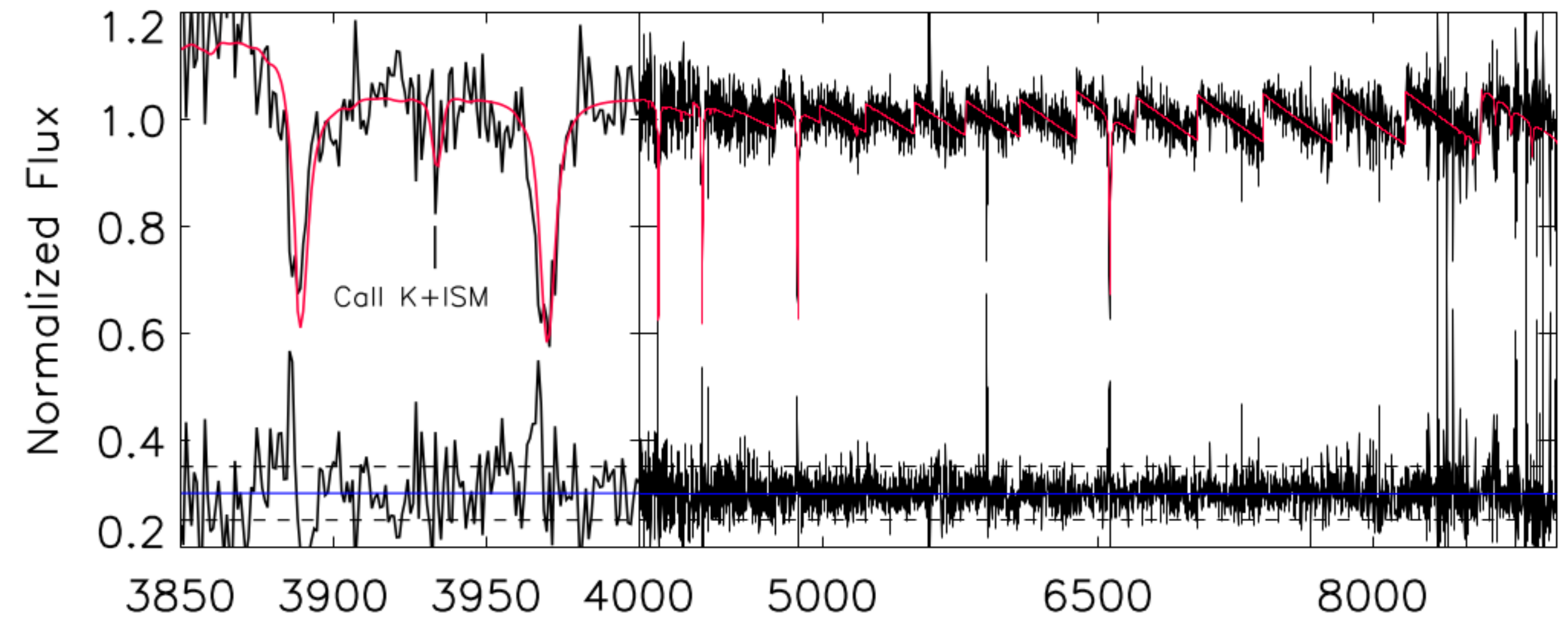}}
{\includegraphics[width=90 mm, angle=0]{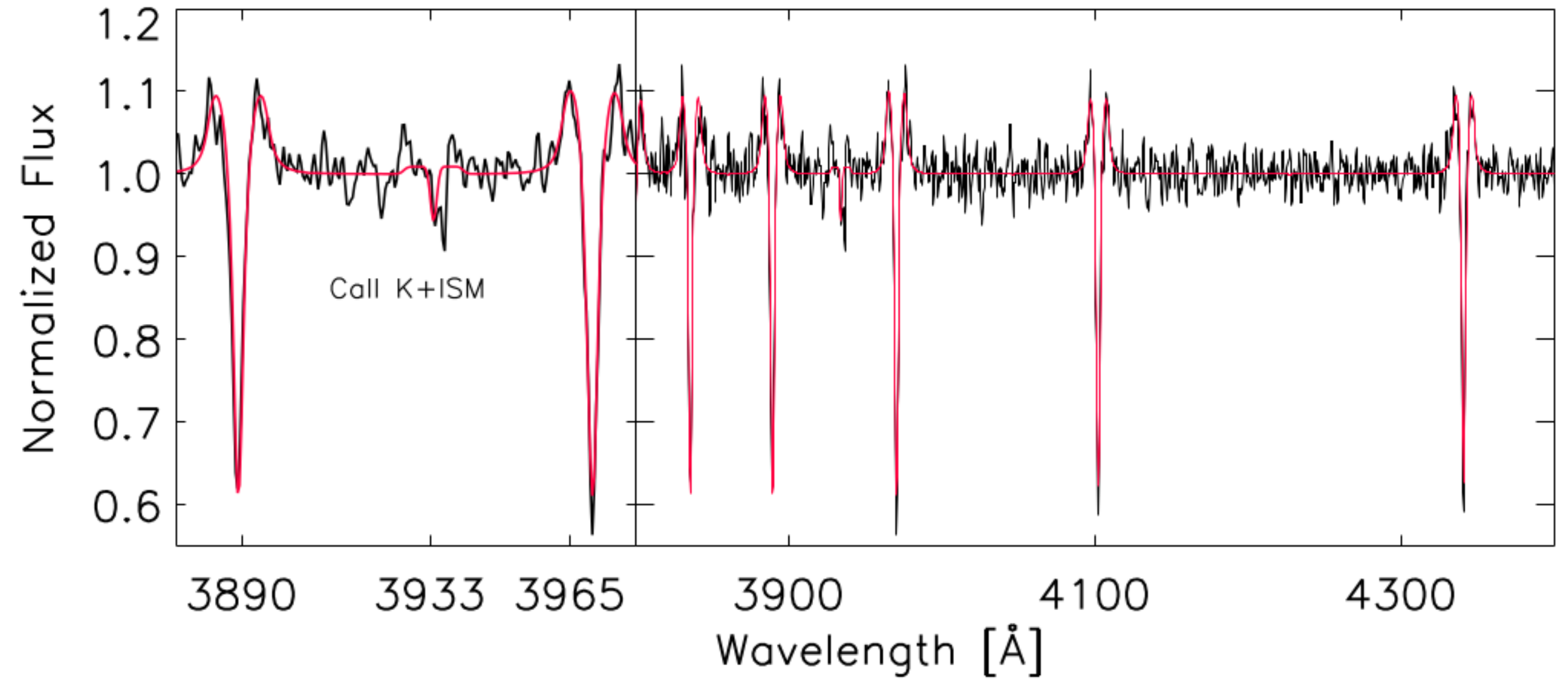}}
\end{center}
\caption{Upper panel: BOSS spectrum of J0023+0207 (black line) and the best fit obtained with FERRE (red line), after dividing the spectra in  segments  of about $\sim$300-400\AA\, which are normalized by their average fluxes. The residuals are also shown in the lower part of both panels.\\
Lower panel: OSIRIS spectrum of J0023+0307 (black line) and the best fit obtained with FERRE (red line). Both the observed and synthetic spectra have been normalized with a running-mean filter with a width of 30 pixels.}
\label{boss}
\end{figure}

\section{Observations} \label{obs}

After identifying J0023+0307 in the SDSS/BOSS survey as an extremely metal-poor candidate, we observed it with the Intermediate dispersion Spectrograph and Imaging System (ISIS) on the 4.2m William Herschel Telescope (WHT) in two different observing runs:
the C54 program on 2016 October 30 and 31, and program (17a)C31 on 2017 September 17 and 21. The adopted set-up was R600B and R600R gratings, the GG495 filter in the red arm and the default dichroic (5300 \AA) with a 1 arcsec slit, providing a resolving power of $R \sim 2400$ in the blue arm. A total of 69 exposures of 1800\,s were taken, resulting in a coadded spectrum with a S/N of $\sim$170 pix$^{-1}$ at 4500 \AA. The star SDSS J1029$+$1729 was observed with the same setup for comparison, and details of the observations are provided in \citet{agu17I}. 

In addition, J0023+0307 was observed with the Optical System for Imaging and low-intermediate-Resolution Integrated Spectroscopy (OSIRIS) and Gran Telescopio Canarias (GTC) between 2016 31 December  and 2017 1 January using Director's Discretionary Time (GTC08-16BDDT). Ten individual exposures of 1300\,s were taken and coadded to produce a spectrum with a S/N$\sim$60 pix$^{-1}$ at 4500 \AA.
Both the ISIS and the OSIRIS data were reduced (bias subtraction, flat-fielding and wavelength calibration using CuNe $+$ CuAr lamps) within the \emph{onespec} package in IRAF\footnote{IRAF is distributed by the National Optical Astronomy Observatory, which is operated by the Association of Universities for Research in Astronomy (AURA) under cooperative agreement with the National Science Foundation} \citep{tod93}. For further details see \citet{agu16,agu17I}.
\begin{figure*}
\begin{center}
{\includegraphics[width=180 mm, angle=180]{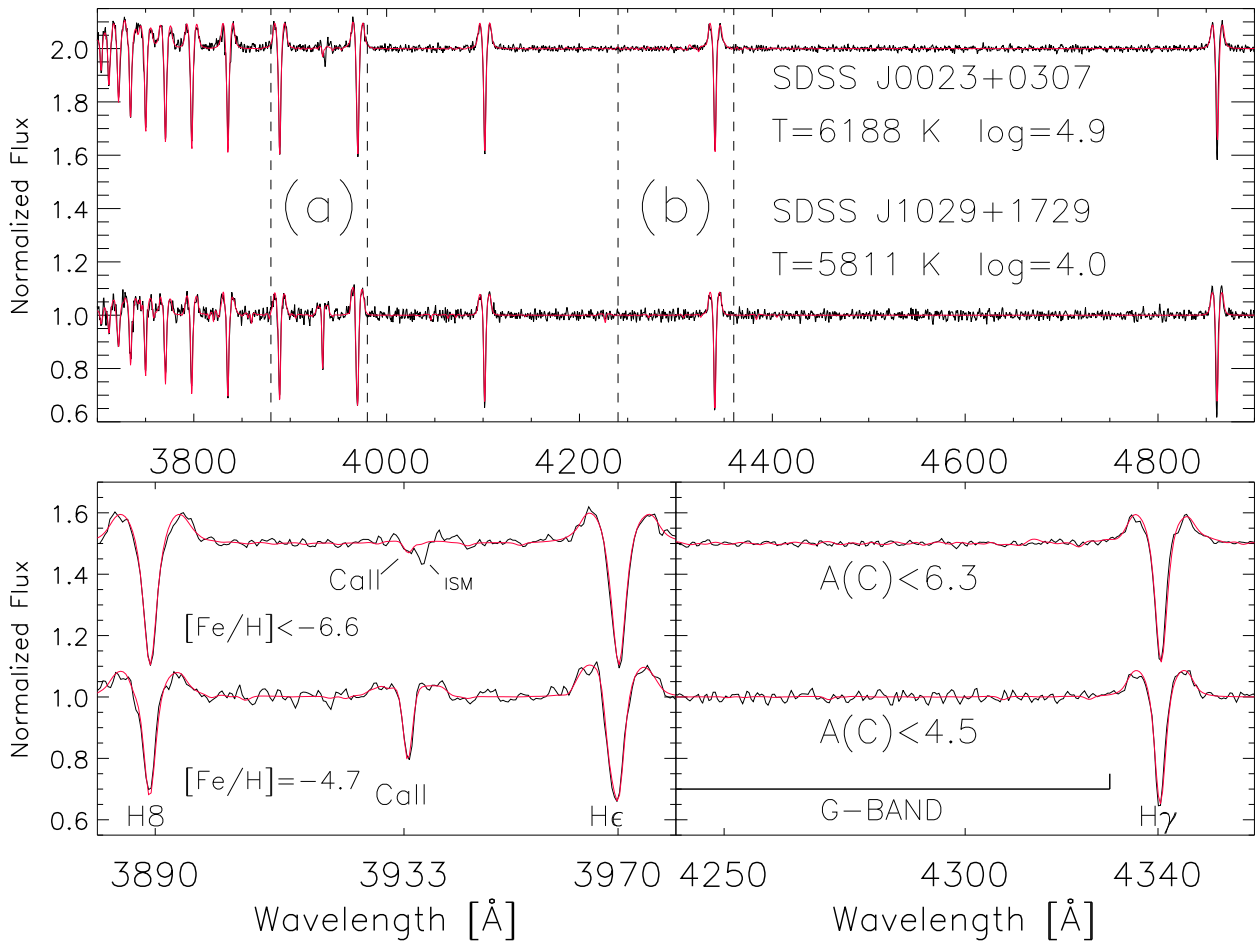}}

\end{center}
\caption{Upper panel: ISIS spectrum of J0023+0307 and J1029+1729 (black line) and the best fits obtained with FERRE (red line). The bottom panels show a details of the Ca II H \& K region (a) and the G-band (b), respectively. Main stellar parameters are shown.
}
\label{isis}
\end{figure*}

\section{Analysis}\label{anali}
The analysis of J0023+0307 was performed in the same way as explained in \citet{alle14} and \citet{agu18I}. We computed a set of model atmospheres using Kurucz's ATLAS9, as in \citet{mez12}, and a grid of synthetic spectra  with the ASS$\epsilon$T code \citep{koe08}, as in \citet{agu17I}. The micro-turbulence and $\alpha$-element abundance  were fixed at 2\,km s$^{-1}$ and $[\alpha/\rm Fe]$=0.4, respectively. We provide further details below.

\subsection{SDSS Analysis}\label{sloananalisis}
 Our first analysis of the BOSS spectrum of J0023+0307 (R.A.$=00^{h} 23^{m} 14^{s}.00$, DEC.=$+03^{0} 07^{'} 58^{''}.07$ (J2000), mag(g)=17.90$\pm 0.01$ and $v_{rad}= -110\pm 9$\,km s$^{-1}$) with FERRE\footnote{FERRE is available from github.com/callendeprieto/ferre}  led to the parameters $T_{\rm eff}=6295 \pm 36$\,K,
$\logg=4.7 \pm 0.2$, [Fe/H]$=-3.9 \pm 0.8$   and [C/Fe]$\sim+0.3$, where the quoted uncertainties are internal estimates provided by FERRE from  inverting the curvature matrix. In the case of [C/Fe], the code derives an extremely large uncertainty, indicating this quantity remains largely unconstrained from these data. Figure \ref{boss} shows the original SDSS/BOSS spectrum and the best fitting. By splitting the spectrum in a number of sections and dividing the flux in each by its mean value we preserve information on the continuum slope, erasing errors in the flux calibration and/or the effect of interstellar extinction. The modest S/N in the blue part of the spectrum does not allow us to derive a reliable metallicity,  but we selected the star as a promising metal-poor candidate for follow-up. We note that using machine learning techniques, \citet{mil15} proposed J0023+0307 as a metal-poor star with [Fe/H]$\sim-2.5$.
Using machine learning techniques, \citet{mil15} proposed J0023+0307 as a metal-poor star with [Fe/H]$\sim-2.5$.

\subsection{ISIS Analysis}\label{analisis}
We simultaneously derive the effective temperature, surface gravity, metallicity and carbon abundance using FERRE (see \citet{alle06} and \citet{agu17I}. A running-mean filter with a window of 30 nearby pixels was used for continuum normalization, since the observations were not provide a flux-calibrated. Fig.~\ref{isis} illustrates our analysis of J0023+0307, and  compare it with that of the most metal-poor star known,  J1029+1729,  identified by \citet{caff11}. For J0023+0307 we derive $T_{\rm eff}=$6188$\pm 84$\,K ($\sim 350$ K warmer than J1029+1729). In addition, our estimated gravity, \logg=4.9$\pm0.5$, also classifies J0023+0307 as dwarf star.

The calcium interstellar medium contribution (ISM) absorption is stronger than the stellar Ca II K line at 3933 \AA, which is barely detected and might include further blending with overlapping ISM contributions. In the absence of any other metallic line in the entire ISIS spectrum, we are able to set an upper-limit of [Fe/H]$<-6.6$, well below the 1D-LTE metallicity value of [Fe/H]$=-4.7$ for J1029+1729 \citep{caff12I}.
The G-band is not visible in the J0023+0307 ISIS spectrum, as it is the case for J1029+1729. 

Since no CH features are detected in the G-Band, we are only able to provide a conservative upper limit of A(C)$<6.3$ (see Fig. \ref{isis}, panel b).  We have also used Markov Chain Monte Carlo analysis to determine the probability distribution for the carbon abundance, illustrated in Fig. 3 (upper-panel), which is consistent with the eyeballed upper limit mentioned above. However, the most likely value could be at [C/Fe]$\lesssim$+2.0 (A(C)$\lesssim$3.8), as shown in the figure.
  Recently, thanks to a very high quality HIRES spectrum (S/N$\sim 700$ and $R\sim 95,000$), \citet{pla16} have measured the carbon abundance of the well-known extremely metal-poor stars 
G64-12 and G64-37 at [C/Fe]$=+1.1$ even though these
warm stars were considered for decades as carbon-normal metal-poor stars.
Higher resolution and/or signal-to-noise data will help setting 
more stringent limits to the metallicity and carbon abundance 
of J0023+0307.

\subsection{OSIRIS analysis}\label{analosiris}
\begin{figure}\label{famo}
\begin{center}
{\includegraphics[width=90 mm]{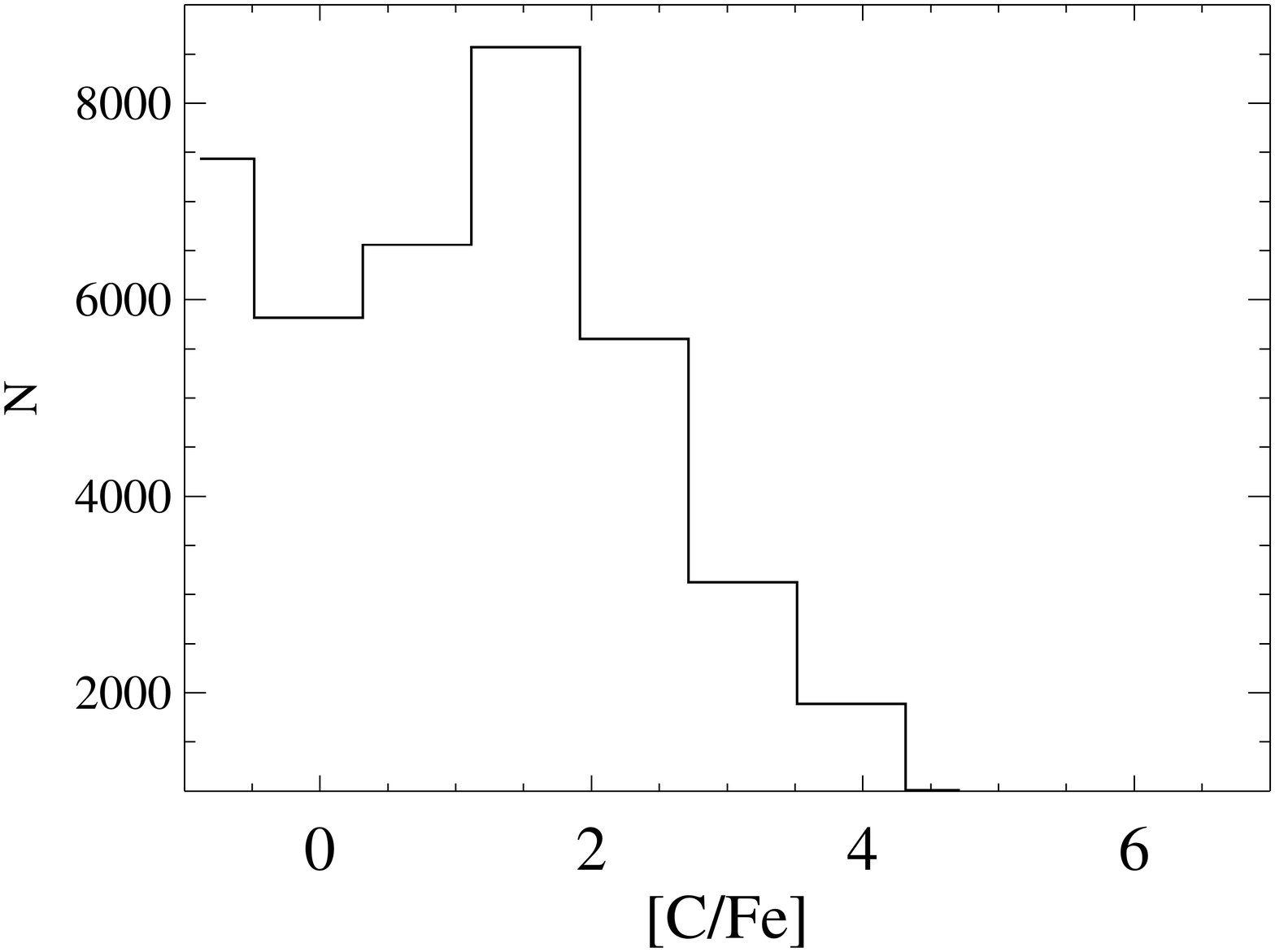}}
{\includegraphics[width=90 mm]{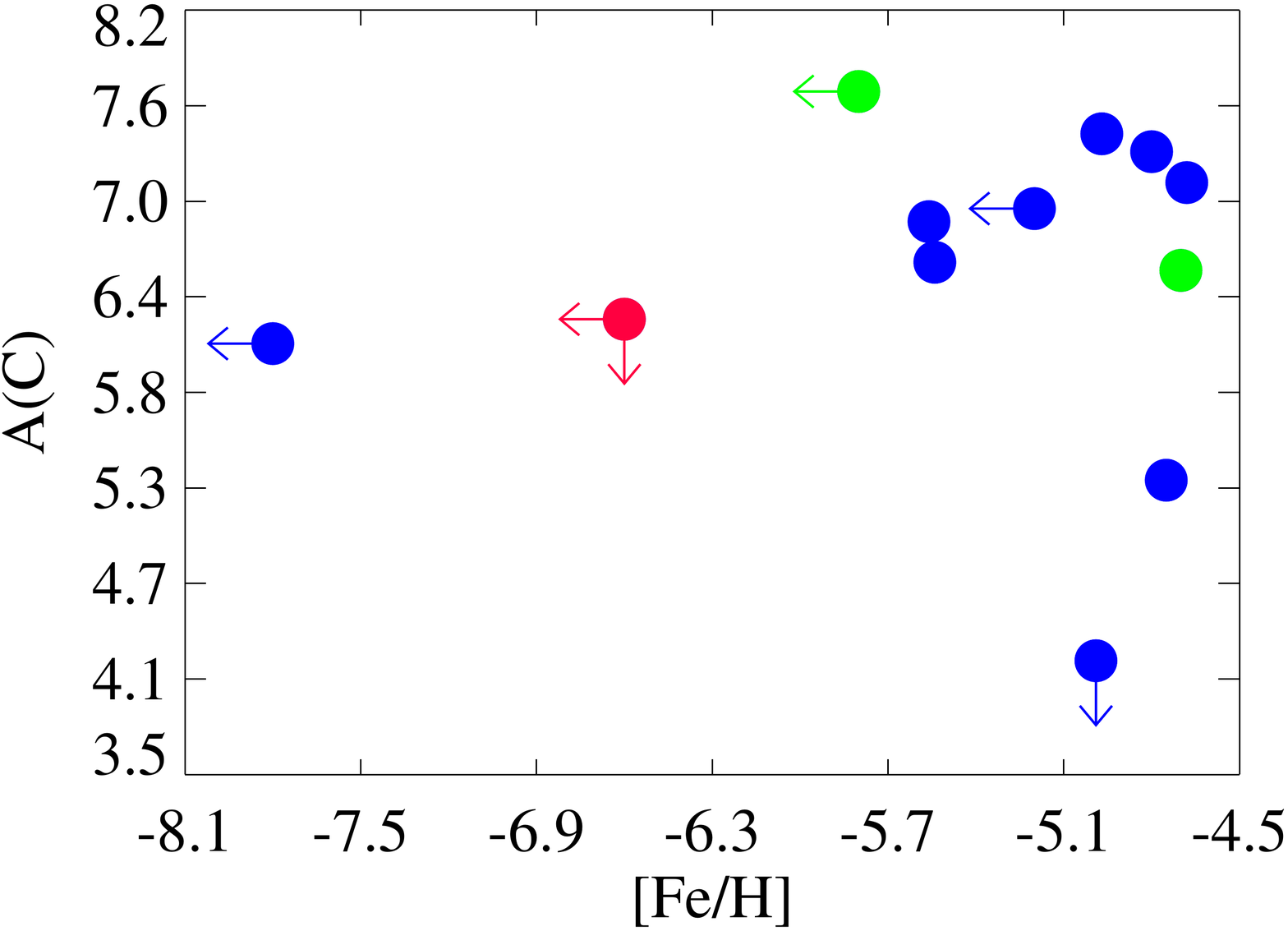}}
\end{center}
\caption{{\it Upper panel}: The number of Markov Chain Monte Carlo experiments vs the carbon-to-iron ratio computed with FERRE.\\
{\it Lower panel}: Carbon abundance vs. iron abundance in 1D-LTE in the ten stars in the [Fe/H]$<$-4.5 regime already in the literature: \citet{chris04,fre05,nor07,caff11,han14,bes15,boni18}, including J0815+4729 \citep{agu18I} and J1313--0019 \citep{alle15,fre15,agu17I}, shown as green circles, and
J0023+0307 (red circle).}
\end{figure}

\ref{famo}
Since the OSIRIS spectrum has significantly lower S/N than the ISIS data, and no extra information about the carbon abundance can be extracted from it, we simply perform an independent analysis along the same lines described in \S \ref{isis}. From this analysis we arrive at $T_{\rm eff}=6140 \pm 132$\,K; $\logg=4.8 \pm0.6$; [C/Fe]$< +3.9$; and  [Fe/H]$<-6.0$. This result supports the analysis above as shown in the lower panel of Fig.~\ref{boss}.

\section{Discussion and conclusions}\label{conclusion}

The star we present and analyze in this letter, J$0023+0307$ is one of the two most iron-poor known to date. The data in hand can only set upper limits for the iron and carbon abundances, but the latter appears to be lower than the typical value of A(C) $\sim7.0$ exhibited by most stars at [Fe/H]$<-4.5$.
From a theoretical perspective, the formation of extremely iron-poor (with [Fe/H]$\lesssim -4.5$) low-mass stars with normal carbon abundance ([C/Fe]$\lesssim +0.7$) remains a open question \citep[see e.g.][]{brom13}.
\citet{brom03} suggested a minimum metallicity is required to cool the collapsing cloud to form low-mass stars (with $M \lesssim 0.8-1 M_{\odot}$) in the Early Universe. Due to the high carbon abundance, the overall metal mass fraction of SMSS J$0313-6708$ ([Fe/H]$<-7.1$) is $\log (Z/Z_{\odot}) \sim -2.1$, which would not violate the theoretical limit for cloud fragmentation driven by radiative cooling by metal lines. On other other hand J$1029+1729$, with [C/Fe]$<0.7$, has $\log (Z/Z_{\odot}) \sim -5$, and favors other fragmentation mechanisms, such as dust cooling \citep{sche12,chi17} or turbulent fragmentation \citep{gre12}.
 In the A(C)$-$[Fe/H] diagram \citep{spi13,yoo16}, this star could fit nicely in the Group III region. Interestingly enough, the authors proposed that all stars in to this group should be CEMP stars.
The  identification of J0023+0307 may push down the metallicity threshold for the formation of low-mass stars. This new limit could be as low as $\log (Z/Z_{\odot}) < -6.6$ if the carbon-to-iron ratio of the star is solar, and even lower if the Ca K line we measure overlaps with a substantial interstellar contribution. Newer observations with higher spectral resolution and very high signal-to-noise ratio are required and can only be obtained with the largest telescopes in the world.

\begin{acknowledgements}
DA acknowledges the Spanish Ministry of Economy and Competitiveness 
(MINECO) for the financial support received in the form of a 
Severo-Ochoa PhD fellowship, within the Severo-Ochoa International PhD 
Program.
DA, CAP, JIGH, and RR also acknowledge the Spanish ministry project MINECO AYA2014-56359-P. JIGH acknowledges financial support from the Spanish Ministry of Economy and Competitiveness (MINECO) under the 2013 Ram\'on y Cajal program MINECO RYC-2013-14875. The authors thankfully acknowledge the technical expertise and assistance provided by the Spanish Supercomputing Network (Red Espanola de Supercomputaci\'on) and Antonio Dorta in particular, as well as the computer resources used: the LaPalma Supercomputer, located at the Instituto de Astrof\'isica de Canarias. This paper is based on observations made with the Gran Telescopio de Canarias (GTC) and with the William Herschel Telescope, operated by the Isaac Newton Group at the Observatorio del  Roque de los Muchachos, La Palma, Spain, of the Instituto de Astrof{\'i}sica de Canarias. We would like to thank GRANTECAN and ING staff members for their efficiency during the observing runs.\\
\end{acknowledgements}
\bibliography{biblio}

\end{document}